\documentclass{article} 

\usepackage[english]{babel} 
\usepackage{amssymb}
\usepackage{amsmath}
\usepackage{txfonts}
\usepackage{mathdots}
\usepackage[classicReIm]{kpfonts}
\usepackage[pdftex]{graphicx}
\usepackage{geometry} 
\begin{document}

\title{%
  Searching of interesting itemsets for negative association rules 
  }

\author{Hyeok Kong, Dokjun An,  Douk Han \\
	Faculty of Mathematics, Kim Il Sung University, Pyongyang, D. P. R. K. \\
	hyeok\_kong@yahoo.com\\
	matadj79@yahoo.com\\
	 hdu0322@yahoo.com
	}

\maketitle

\noindent \textbf{Abstract:}  In this paper, we propose an algorithm of searching for both positive and negative itemsets of interest which should be given at the first stage for positive and negative association rules mining. Traditional association rule mining algorithms extract positive association rules based on frequent itemsets, for which the frequent itemsets, i.e. only positive itemsets of interest are searched. Further, there are useful itemsets among the frequent itemsets pruned from the traditional algorithms to reduce the search space, for mining of negative association rules. Therefore, the traditional algorithms have not come true to find negative itemsets needed in mining of negative association rules. Our new algorithm to search for both positive and negative itemsets of interest prepares preconditions for mining of all positive and negative association rules.   


\section{Related works}

\noindent We gave the definitions of positive and negative itemsets of interest in \cite{Kong2018}. Apriori, a traditional association rule mining algorithm \cite{Agrawal1994}, has not come true to mine negative association rules because it pruned certain frequent itemsets among the candidate itemsets(C${}_{k}$), not included in the prior family of frequent itemsets(L${}_{k}$${}_{-1}$). 

\noindent Also, the search space problem arised because of a great number of positive and negative itemsets, so it was important to identify only itemsets of interest needed for mining of association rules.${}^{ }$

\noindent In \cite{Kong2016,Almasi2015}, rare association rule mining such as $\mathrm{\neg}$A$\mathrm{\to}$$\mathrm{\neg}$C was studied, but it was only a kind of negative association rules, so they did not give clear definition and consider searching problem dealing with all possible cases of negative association rules. 

\noindent The definitions of itemsets of interest are as follows.\cite{Kong2018}

\noindent \textbf{Definition 1.} X$\mathrm{\to}$Y is called \textbf{a positive association rule of interest}, and X$\mathrm{\cup}$Y is called \textbf{a positive itemset of interest}, if they satisfy the following conditions.

\begin{enumerate}
\item  $X\bigcap Y=\phi $

\item  sprt(X$\mathrm{\cup}$Y) $\mathrm{\ge}$ minsprt ,  

\item  {\textbar}sprt(X$\mathrm{\cup}$Y) $-$ sprt(X)sprt(Y){\textbar} $\mathrm{\ge}$ mininterest,

\item  sprt(X$\mathrm{\cup}$Y)/sprt(X) $\mathrm{\ge}$ minconf
\end{enumerate}

\noindent Otherwise, if {\textbar}sprt(X$\mathrm{\cup}$Y)$-$sprt(X)sprt(Y){\textbar} $<$ mininterest, or sprt(X$\mathrm{\cup}$Y)/sprt(X) $<$ minconf, then the rule X$\mathrm{\to}$Y is not of interest, and X$\mathrm{\cup}$Y is called \textbf{an uninteresting itemset}.

\noindent Conversely, if Q is a positive itemset of interest, there is at least one expression Q= X$\mathrm{\cup}$Y such that X and Y satisfy the above 4 conditions for positive association rules of interest.

\noindent \textbf{Definition 2.}  The rule A$\mathrm{\to}$$\mathrm{\neg}$B is called \textbf{a negative association rule of interest} and A$\mathrm{\cup}$B is calle \textbf{a negative itemset of interest} if they satisfy the following conditions.

\begin{enumerate}
\item  $A\bigcap B=\phi ,$

\item  sprt(A) $\mathrm{\ge}$ minsprt , sprt(B) $\mathrm{\ge}$ minsprt , sprt(A$\mathrm{\cup}$$\mathrm{\neg}$B) $\mathrm{\ge}$ minsprt,

\item  sprt(A$\mathrm{\cup}$$\mathrm{\neg}$B) $-$ sprt(A)sprt($\mathrm{\neg}$B) $\mathrm{\ge}$ mininterest,

\item  sprt(A$\mathrm{\cup}$$\mathrm{\neg}$B)/sprt(A) $\mathrm{\ge}$ minconf.
\end{enumerate}

\noindent Otherwise, the rule A$\mathrm{\to}$$\mathrm{\neg}$B is not of interest, and A$\mathrm{\cup}$B is \textbf{an uninteresting itemset}. On the other hand, if Q is a negative itemset of interest, there is at least one expression Q=A$\mathrm{\cup}$B such that one of the rules: A$\mathrm{\to}$$\mathrm{\neg}$B, or $\mathrm{\neg}$A$\mathrm{\to}$B, or $\mathrm{\neg}$A$\mathrm{\to}$$\mathrm{\neg}$B, is a valid negative association rule of interest.

\noindent Thus, uninteresting itemsets are any itemsets in a database which exclude both positive and negative itemsets of interest. These itemsets need to be pruned to reduce the space searched in mining.

\noindent In other words, there are a large number of infrequent itemsets related to uninteresting association rules. If we could extract only positive and negative itemsets of iterest among so many itemsets, the space searched would be extremely reduced.  

\section{Design of algorithm}

\noindent For the purpose to develop an algorithm of mining both positive and negative association rules, we first consider data for the algorithm. Database \textit{D}, minimum support \textit{minsprt}, minimum confidence \textit{minconf} and mimimum interest \textit{mininterest} are given in the algorithm. The resulting data of the algorithm are PS, a family of positive itemsets of interest, and NS, a family of negative ones.

\noindent During the running of the algorithm, Freq${}_{k}$, a family of frequent itemsets must be generated at the \textit{k}th pass of the algorithm, and based on them, the sets of positive and negative itemsets, namely P${}_{k}$ and N${}_{k}$, must be also generated respectively, where P${}_{k}$ is the same as L${}_{k}$, the family of frequent itemsets in the traditional algorithms of mining association rules.

\noindent Now, let Temp${}_{k}$=P${}_{k}$$\mathrm{\cup}$N${}_{k}$. Then, C${}_{k}$, the candidate set in the traditional algorithms of mining association rules, is a subset of Temp${}_{k}$, and every member of it must include at least one subset which is a member of P${}_{k}$${}_{-1}$. Therefore, Temp${}_{k}$ is a family of k-itemsets and each itemset of Temp${}_{k}$ is an union of any two frequent itemsets in Freq${}_{i}$(${\rm 1}\le i\le k-1$). That is, for itemsets A in Freq${}_{i}$${}_{0 }$and B in Freq${}_{i}$${}_{1}$ (${\rm 1}\le i_{{\rm 0}} ,\, \, \, i_{1} \le k-1$), if A$\mathrm{\cup}$B is a k-itemset, A$\mathrm{\cup}$B is appended into Temp${}_{k}$. And each itemset in Temp${}_{k}$ must be counted in the database D.

\noindent Next, if an itemset Q= X$\mathrm{\cup}$Y in P${}_{k}$${}_{ }$satisfies {\textbar}sprt(X$\mathrm{\cup}$Y)$-$sprt(X)sprt(Y){\textbar}$<$mininterest for any X and Y, then Q is the uninteresting frequent itemset, so it must be pruned from P${}_{k}$. P${}_{k}$ with all uninteresting itemsets pruned from it is appended to PS. Similarly, if an itemset Q= X$\mathrm{\cup}$Y in N${}_{k}$${}_{ }$satisfies {\textbar}sprt(X$\mathrm{\cup}$Y)$-$sprt(X)sprt(Y){\textbar}$<$mininterest for any X and Y, then Q is the uninteresting frequent itemset, so it must be pruned from N${}_{k}$. N${}_{k}$ with all uninteresting itemsets pruned from it is appended to NS. 

\noindent The end conditions of the above loop are P${}_{k}$$\ne \emptyset $and N${}_{k}$$\ne \emptyset $. PS and NS are output as results of the algorithm.

\noindent The algorithm is as follows.

\noindent \textbf{[Algorithm] Searching of Interesting Itemsets}

\noindent Input data: D(database), minsprt(minimum support), minconf(minimum confidence), mininterest(minimum interest) 

\noindent Output data: PS(family of positive itemsets of interest),  NS(family of negative itemsets of interest )

\begin{enumerate}
\item PS=$\mathrm{\emptyset}$; NS=$\mathrm{\emptyset}$;

\item Freq${}_{1}$=family of 1-frequent itemsets; // First pass of D

k=1;

\noindent // Generation of all positive and negative k-itemsets of interest

\item do $\{$ 

     k++;

// Generation of possible k-itemsets

\noindent Temp${}_{k}$=$\{$A$\mathrm{\cup}$B{\textbar} A$\mathrm{\in}$Freq${}_{i0}$, B$\mathrm{\in}$ Freq${}_{i1}$, (${\rm 1}\le i_{{\rm 0}} ,\, \, \, i_{1} \le k-1$), {\textbar}A$\mathrm{\cup}$B{\textbar}=k$\}$; 

  //Counting of k-itemsets in transactions in D

for $\mathrm{\forall}$t $\mathrm{\in}$ D do $\{$

   Temt=$\{$k-itemset{\textbar} k-itemset$\mathrm{\subseteq}$t, k-itemset$\mathrm{\in}$Temp${}_{k}$$\}$;

   for $\mathrm{\forall}$itemset $\mathrm{\in}$Temt do 

    itemset.count= itemset.count+1;

$\}$

  // Generation of positive and negative k-itemsets based on k-candidate itemsets and k-frequent itemsets

C${}_{k}$=$\{$ k-itemset{\textbar} k-itemset$\mathrm{\in}$Temp${}_{k}$, $\mathrm{\exists}$Sub$\mathrm{\subset}$k-itemset: Sub$\mathrm{\in}$P${}_{k}$${}_{-1}$$\}$; 

  Freq${}_{k}$=$\{$c{\textbar} c$\mathrm{\in}$C${}_{k}$, sprt(c)=c.count/{\textbar}D{\textbar} $>$= minsprt$\}$; 

P${}_{k}$= Freq${}_{k}$; // k-positive itemsets

NN${}_{k}$= Temp${}_{k}$- P${}_{k}$ 

// Pruning of uninteresting k-positive itemsets

for $\mathrm{\forall}$itemset $\mathrm{\in}$P${}_{k}$ do

   if itemset: uninteresting itemset then  P${}_{k}$=P${}_{k}$ -- $\{$itemset$\}$;

  PS=PS$\mathrm{\cup}$P${}_{k}$; 

 //Pruning of uninteresting k-negative itemsets

N${}_{ k}$= $\{$itemset $\mathrm{\in}$NN${}_{ k}$, itemset: negative itemset$\}$; // k-negative itemset

for $\mathrm{\forall}$itemset $\mathrm{\in}$N${}_{k}$ do

\noindent if itemset: uninteresting itemset then  N${}_{k}$=N${}_{k}$ -- $\{$itemset$\}$;

\noindent NS=NS$\mathrm{\cup}$N${}_{k}$;

  $\}$ while (P${}_{k}$${}_{-1}$$\ne \Phi ,$ N${}_{k}$${}_{-1}$$\ne \Phi $);

 // Reseult output

 \item output PS, NS;
\end{enumerate}

\noindent When uninteresting itemsets are pruned from the space searched of exponential size using the above algorithm, the space searched is extremely reduced.

\section{Application of algorithm}

\noindent The following example shows an application of the algorithm. A transaction database of 10 transactions is shown in Table 1. There are 6 items A, B, C, D, E and F. And, minsprt=0.3 and mininterest=0.07 are supposed.

\noindent Table 1. Example database

\begin{tabular}{|p{0.7in}|p{0.7in}|} \hline 
Transaction ID & Items \\ \hline 
T1 & A, B, D \\ \hline 
T2 & A, B, C, D \\ \hline 
T3 & B, D \\ \hline 
T4 & B, C, D, E \\ \hline 
T5 & A, C, E \\ \hline 
T6 & B, D, F \\ \hline 
T7 & A, E, F \\ \hline 
T8 & C, F \\ \hline 
T9 & B, C, F \\ \hline 
T10 & A, B, C, D, F \\ \hline 
\end{tabular}

Through the first scan of database, the support of each item is counted as follows.

\noindent A: 5/10=0.5

\noindent B: 7/10=0.7

\noindent C: 6/10=0.6

\noindent D: 6/10=0.6

\noindent E: 3/10=0.3

\noindent F: 5/10=0.5

\noindent From minsprt=0.3, all items are 1-frequent itemsets. So Freq${}_{1}$=$\{$A,B,C,D,E,F$\}$. Temp${}_{2}$=$\{$AB, AC, AD, AE, AF, BC, BD, BE, BF, CD, CE, CF, DE, DF, EF$\}$ is generated from Freq${}_{1}$, and Freq${}_{2}$=P${}_{2}$=$\{$AB, AC, AD, BC, BD, BF, CD, CF$\}$ is generated through Temt and C${}_{2}$, based on${}_{ }$minsprt=0.3. So NN${}_{2}$= Temp${}_{2}$-P${}_{2}$=$\{$AE, AF, BE, CE, DE, DF, EF$\}$.

\noindent For pruning of uninteresting itemsets from P${}_{2}$ on the basis of mininterest=0.07, the interest measure of each itemset is calculated as follows.

\noindent {\textbar}sprt(A$\mathrm{\cup}$B)$-$sprt(A)sprt(B){\textbar}=0.05$<$mininterest,

\noindent {\textbar}sprt(A$\mathrm{\cup}$C)$-$sprt(A)sprt(C){\textbar}=0$<$mininterest, 

\noindent {\textbar}sprt(A$\mathrm{\cup}$D)$-$sprt(A)sprt(D){\textbar} = 0 $<$ mininterest,

\noindent {\textbar}sprt(B$\mathrm{\cup}$C)$-$sprt(B)sprt(C){\textbar}=0.02$<$mininterest,

\noindent {\textbar}sprt(B$\mathrm{\cup}$D)$-$sprt(B)sprt(D{\textbar}=0.18$>$mininterest,

\noindent {\textbar}sprt(B$\mathrm{\cup}$F)$-$sprt(B)sprt(F){\textbar}=0.05$<$mininterest,

\noindent {\textbar}sprt(C$\mathrm{\cup}$D)$-$sprt(C)sprt(D){\textbar}=0.06$<$mininterest,

\noindent {\textbar}sprt(C$\mathrm{\cup}$F)$-$sprt(C)sprt(F){\textbar}=0$<$mininterest.

\noindent As shown in the above calculations, only B$\mathrm{\cup}$D is an itemset of interest and the rest of P${}_{2}$ are all uninteresting itemsets. From this, pruning all uninteresting itemsets from P${}_{2}$ before appending them to PS gives P${}_{2}$=$\{$BD$\}$.

\noindent Similarly, the interest measure of each itemset in NN${}_{2}$${}_{ }$is calculated as follows.

\noindent {\textbar}sprt(A$\mathrm{\cup}$E)$-$sprt(A)sprt(E){\textbar}=0.05$<$mininterest,

\noindent {\textbar}sprt(A$\mathrm{\cup}$F)$-$sprt(A)sprt(F){\textbar}=0.05$<$mininterest, 

\noindent {\textbar}sprt(B$\mathrm{\cup}$E)$-$sprt(B)sprt(E){\textbar}=0.11$>$mininterest,

\noindent {\textbar}sprt(C$\mathrm{\cup}$E)$-$sprt(C)sprt(E){\textbar}=0.02$<$mininterest,

\noindent {\textbar}sprt(D$\mathrm{\cup}$E)$-$sprt(D)sprt(E){\textbar}=0.02$<$mininterest,

\noindent {\textbar}sprt(D$\mathrm{\cup}$F)$-$sprt(D)sprt(F){\textbar}=0.1$>$mininterest,

\noindent {\textbar}sprt(E$\mathrm{\cup}$F)$-$sprt(E)sprt(F){\textbar}=0.05$<$mininterest.

\noindent As shown in the above calculations, only B$\mathrm{\cup}$E and D$\mathrm{\cup}$F are itemsets of interest and the rest of N${}_{2}$ are all uninteresting sets. From this, pruning all uninteresting itemsets from N${}_{2}$ before appending them to NS gives N${}_{2}$=$\{$BE, DF$\}$.

\noindent Next, Temp${}_{3}$=$\{$ABC, ABD, ABE, ABF, ACD, ACE, ACF, ADE, ADF, AEF, BCD, BCE, BCF, BDE, BDF, BEF, CDE,CDF, CEF, DEF$\}$ is generated from Freq${}_{1}$ and Freq${}_{2}$, and Freq${}_{3}$=P${}_{3}$=$\{$ABD, BCD$\}$ is generated through Temt and C${}_{3}$, based on${}_{ }$minsprt=0.3. So NN${}_{3}$= Temp${}_{3}$-P${}_{3}$=$\{$ABC, ABE, ABF, ACD, ACE, ACF, ADE, ADF, AEF, BCE, BCF, BDE, BDF, BEF, CDE,CDF, CEF, DEF$\}$.

\noindent For pruning of uninteresting itemsets from P${}_{3}$ on the basis of mininterest=0.07, the interest measure of each itemset is calculated as follows(for the sake of convenience, A$\mathrm{\cup}$B$\mathrm{\cup}$C is represented as ABC, and A$\mathrm{\cup}$B is also represented as AB).

\noindent {\textbar}sprt(ABD)$-$sprt(AB)sprt(D){\textbar}=0.12$>$mininterest,

\noindent {\textbar}sprt(BCD)$-$sprt(B)sprt(CD){\textbar}=0.09$>$mininterest.

\noindent This shows that ABD and BCD are both itemsets of interest. So P${}_{3}$=$\{$ABD, BCD$\}$.

\noindent Similarly, the interest measure of each itemset in NN${}_{3 }$is calculated as follows.

\noindent {\textbar}sprt(ABE)$-$sprt(AB)sprt(E){\textbar}=0.09$>$mininterest,

\noindent {\textbar}sprt(ADE)$-$sprt(AD)sprt(E){\textbar}=0.09$>$mininterest,

\noindent {\textbar}sprt(BDE)$-$sprt(BD)sprt(E){\textbar}=0.08$>$mininterest,

\noindent {\textbar}sprt(BDF)$-$sprt(BD)sprt(F){\textbar}=0.1$>$mininterest,

\noindent {\textbar}sprt(BEF)$-$sprt(BF)sprt(E){\textbar}=0.09$>$mininterest,

\noindent {\textbar}sprt(CDF)$-$sprt(C)sprt(DF){\textbar}=0.08$>$mininterest,

\noindent {\textbar}sprt(CEF)$-$sprt(CE)sprt(F){\textbar}=0.1$>$mininterest.

\noindent From the above calculations, pruning all uninteresting itemsets from N${}_{3}$ before appending them to NS results in N${}_{3}$=$\{$ABE, ADE, BDE, BDF, BEF, CDF, CEF$\}$.

\noindent Next, Temp${}_{4}$=$\{$ABCD, ABCF, ABDE, ABDF, BCDE, BCDF$\}$ is generated from Freq${}_{1}$, Freq${}_{2 }$and Freq${}_{3}$, and Freq${}_{4}$=P${}_{4}$=$\mathrm{\emptyset}$ is generated through Temt and C${}_{4}$, based on${}_{ }$minsprt=0.3. So NN${}_{4}$= Temp${}_{4}$-P${}_{4}$=$\{$ABCD, ABCF, ABDE, ABDF, BCDE, BCDF$\}$.

\noindent For pruning of uninteresting itemsets from NN${}_{4}$ on the basis of mininterest=0.07, the interest measure of each itemset is calculated as follows.

\noindent {\textbar}sprt(ABCD)$-$sprt(AB)sprt(CD){\textbar}=0.11$>$mininterest,

\noindent {\textbar}sprt(ABDE)$-$sprt(AB)sprt(DE){\textbar}=0.07=mininterest.

\noindent From the above calculations, pruning all uninteresting itemsets from N${}_{4}$ before appending them to NS yields N${}_{4}$=$\{$ABCD, ABDE$\}$.

\noindent Finally, Temp${}_{5}$=$\{$ABCDF$\}$ is generated and Freq${}_{5}$=P${}_{5}$=$\mathrm{\emptyset}$ based on minsprt=0.3. So NN${}_{5}$= Temp${}_{5}$-P${}_{5}$=$\{$ABCDF$\}$. ABCDF is not interesting from calculation of the interest measure. Therefore, pruning the uninteresting itemset from NN${}_{5}$ gives N${}_{5}$=$\mathrm{\emptyset}$. At this point, the loop is finished and the algorithm outputs the following results.

\noindent PS=$\{$P${}_{2}$, P${}_{3}$, P${}_{4}$, P${}_{5}$$\}$= $\{$BD, ABD, BCD$\}$

\noindent NS=$\{$N${}_{2}$, N${}_{3}$, N${}_{4}$, N${}_{5}$$\}$= $\{$BE, DF, ABE, ADE, BDE, BDF, BEF, CDF, CEF, ABCD, ABDE$\}$

\noindent From the application of the algorithm, only 4 positive itemsets of interest among 11 frequent itemsets(excepting 1-itemsets) are obtained, and only 11 negative itemsets of interest among 31 negative itemsets are obtained.

\noindent In this algorithm, it is not considered whether confidences of the rules are greater than minconf or not, identifying the positive and the negative itemsets of interest. It means that each of positive and negative itemsets of interest should be dealt with together with the forth condition in Definitions 1 and 2. Dealing with this condition, the space searched would be more reduced.

\noindent Through the application of the algorithm, we can see that if a frequent itemset is pruned straight away when frequent itemsets of interest are searched, it can be found no longer in the frequent itemsets, but conversely it can be found in infrequent itemsets of interest. After then, if the itemset is pruned when infrequent itemsets of interest are searched, it is removed from the rest of the searched space. If an infrequent itemset is pruned when infrequent itemsets of interest are searched, it does not impact on searching for frequent itemsets of interest.

\section{Conclusion}

\noindent For mining both positive and negative association rules, not only frequent itemsets but also infrequent itemsets should be identified. Identifying frequent itemsets is exactly searching of the space of exponential size of possible items and itemsets, and the number of possible infrequent itemsets is much greater than the number of frequent itemsets. That is, the amount of possible positive and negative itemsets becomes almost double as much.

\noindent In this paper, we proposed an algorithm of searching for both positive and negative itemsets of interest that should be given at the first stage for positive and negative association rules mining. This algorithm prepares preconditions for mining of all positive and negative association rules. And we showed by an example that the space searched is extremely reduced dealing with itemsets of interest.

\noindent \textbf{}

\bibliographystyle{plain}
\bibliography{reference}

\end{document}